\documentclass[twoside,twocolumn,9pt]{article}
\usepackage{extsizes}
\usepackage[super,sort&compress,comma]{natbib} 
\usepackage[version=3]{mhchem}
\usepackage[left=1.5cm, right=1.5cm, top=1.785cm, bottom=2.0cm]{geometry}
\usepackage{balance}
\usepackage{times,mathptmx}
\usepackage{sectsty}
\usepackage{graphicx} 
\usepackage{lastpage}
\usepackage[format=plain,justification=justified,singlelinecheck=false,font={stretch=1.125,small,sf},labelfont=bf,labelsep=space]{caption}
\usepackage{float}
\usepackage{fancyhdr}
\usepackage{fnpos}
\usepackage[english]{babel}
\addto{\captionsenglish}{%
  
}
\usepackage{array}
\usepackage{droidsans}
\usepackage{charter}
\usepackage[T1]{fontenc}
\usepackage[usenames,dvipsnames]{xcolor}
\usepackage{setspace}
\usepackage[compact]{titlesec}
\usepackage{hyperref}
\usepackage{epstopdf}%This line makes .eps figures into .pdf - please comment out if not required.
\newcommand{\etal}{\textit{et~al.}}
\definecolor{cream}{RGB}{222,217,201}

\begin{document}

\pagestyle{fancy}
\thispagestyle{plain}
\fancypagestyle{plain}{

%%%HEADER%%%

\renewcommand{\headrulewidth}{0pt}
}
%%%END OF HEADER%%%

%%%PAGE SETUP - Please do not change any commands within this section%%%
\makeFNbottom
\makeatletter
\renewcommand\LARGE{\@setfontsize\LARGE{15pt}{17}}
\renewcommand\Large{\@setfontsize\Large{12pt}{14}}
\renewcommand\large{\@setfontsize\large{10pt}{12}}
\renewcommand\footnotesize{\@setfontsize\footnotesize{7pt}{10}}
\makeatother

\renewcommand{\thefootnote}{\fnsymbol{footnote}}
\renewcommand\footnoterule{\vspace*{1pt}% 
\color{cream}\hrule width 3.5in height 0.4pt \color{black}\vspace*{5pt}} 
\setcounter{secnumdepth}{5}

\makeatletter 
\renewcommand\@biblabel[1]{#1}            
\renewcommand\@makefntext[1]% 
{\noindent\makebox[0pt][r]{\@thefnmark\,}#1}
\makeatother 
\renewcommand{\figurename}{\small{Fig.}~}
\sectionfont{\sffamily\Large}
\subsectionfont{\normalsize}
\subsubsectionfont{\bf}
\setstretch{1.125} %In particular, please do not alter this line.
\setlength{\skip\footins}{0.8cm}
\setlength{\footnotesep}{0.25cm}
\setlength{\jot}{10pt}
\titlespacing*{\section}{0pt}{4pt}{4pt}
\titlespacing*{\subsection}{0pt}{15pt}{1pt}
%%%END OF PAGE SETUP%%%

%%%FOOTER%%%
\fancyfoot{}

\fancyfoot[RO]{\footnotesize{\sffamily{1--\pageref{LastPage} ~\textbar  \hspace{2pt}\thepage}}}
\fancyfoot[LE]{\footnotesize{\sffamily{\thepage~\textbar\hspace{2pt} 1--\pageref{LastPage}}}}
\fancyhead{}
\renewcommand{\headrulewidth}{0pt} 
\renewcommand{\footrulewidth}{0pt}
\setlength{\arrayrulewidth}{1pt}
\setlength{\columnsep}{6.5mm}
\setlength\bibsep{1pt}
%%%END OF FOOTER%%%

%%%FIGURE SETUP - please do not change any commands within this section%%%
\makeatletter 
\newlength{\figrulesep} 
\setlength{\figrulesep}{0.5\textfloatsep} 

\newcommand{\topfigrule}{\vspace*{-1pt}% 
\noindent{\color{cream}\rule[-\figrulesep]{\columnwidth}{1.5pt}} }

\newcommand{\botfigrule}{\vspace*{-2pt}% 
\noindent{\color{cream}\rule[\figrulesep]{\columnwidth}{1.5pt}} }

\newcommand{\dblfigrule}{\vspace*{-1pt}% 
\noindent{\color{cream}\rule[-\figrulesep]{\textwidth}{1.5pt}} }

\makeatother
%%%END OF FIGURE SETUP%%%

%%%TITLE, AUTHORS AND ABSTRACT%%%
\twocolumn[
  \begin{@twocolumnfalse}
\vspace{0cm}
\sffamily
\begin{tabular}{m{2cm} p{14cm} }

 & \noindent\LARGE{\textbf{Cross-sectional focusing of red blood cells in a constricted microfluidic channel$^\dag$}} \\%Article title goes here instead of the text "This is the title"
\vspace{0.3cm} & \vspace{0.3cm} \\

 & \noindent\large{Asena Abay,\textit{$^{a,b}$}\textit{$^{\ddag}$} Steffen M. Recktenwald,$^{\ast}$\textit{$^{a}$}}\textit{$^{\ddag}$} Thomas John,\textit{$^{a}$} Lars Kaestner,\textit{$^{a,c}$} and Christian Wagner\textit{$^{a}$} \\%Author names go here instead of "Full name", etc.
\vspace{0.3cm} & \vspace{0.3cm} \\

 & \noindent\normalsize{Constrictions in blood vessels and microfluidic devices can dramatically change the spatial distribution of passing cells or particles and are commonly used in biomedical cell sorting applications. However, the three-dimensional nature of cell focusing in the channel cross-section remains poorly investigated. Here, we explore the cross-sectional distribution of living and rigid red blood cells passing a constricted microfluidic channel by tracking individual cells in multiple layers across the channel depth and across the channel width. While cells are homogeneously distributed in the channel cross-section pre-contraction, we observe a strong geometry-induced focusing towards the four channel faces post-contraction. The magnitude of this cross-sectional focusing effect increases with increasing Reynolds number for both living and rigid red blood cells. We discuss how this non-uniform cell distribution downstream of the contraction results in an apparent double-peaked velocity profile in particle image velocimetry analysis and show that trapping of red blood cells in the recirculation zones of the abrupt construction depends on cell deformability.} \\%The abstract goes here instead of the text "The abstract should be..."
\end{tabular}
 \end{@twocolumnfalse} \vspace{1cm}
]
%%%END OF TITLE, AUTHORS AND ABSTRACT%%%

%%%FONT SETUP - please do not change any commands within this section
\renewcommand*\rmdefault{bch}\normalfont\upshape
\rmfamily
\section*{}
\vspace{-1cm}

%%%FOOTNOTES%%%

\footnotetext{\textit{$^{a}$~Dynamics of Fluids, Department of Experimental Physics, Saarland University, Saarbr{\"u}cken, Germany; $^{\ast}$~E-mail: steffen.recktenwald@uni-saarland.de}}
\footnotetext{\textit{$^{b}$~Department of Hematopoiesis, Sanquin Research, Amsterdam, Netherlands.}}
\footnotetext{\textit{$^{c}$~Theoretical Medicine and Biosciences, Saarland University, Homburg, Germany.}}
%Please use \dag to cite the ESI in the main text of the article.
%If you article does not have ESI please remove the the \dag symbol from the title and the footnotetext below.
\footnotetext{\dag~Electronic Supplementary Information (ESI) available: One PDF containing additional experimental data. See DOI: 00.0000/00000000.}
\footnotetext{\ddag~These authors contributed equally to this work.}
%additional addresses can be cited as above using the lower-case letters, c, d, e... If all authors are from the same address, no letter is required

%%%END OF FOOTNOTES%%%

%%%MAIN TEXT%%%%%%%%%%%%%%%%%%%%%%%%%%%%%%%%%%%%%%%%%%%%%%%%%%%%%%%%%%%%%%%%%%%%%%%%%%%%%%%%%%%%%%%%%%%%%%%%%%%%%%%%%%%%%%%%%%
\section{Introduction}
The cardiovascular system is a complex network of branching vessels that transport and distribute blood through our body. Blood is mainly comprised of deformable red blood cells (RBCs) that are suspended in plasma. During laminar blood flow, RBCs preferentially migrate laterally from the vessel walls to the center and generate a cell-free layer near the vessel walls.\cite{Kim2009} This phenomenon, known as the F{\aa}hraeus effect\cite{Fahraeus1929}, determines the unique flow properties of blood in microvessels and is crucial for its physiological functionality.\cite{Secomb2017}

The microscopic flow behavior of RBC suspensions is often studied using straight microfluidic channels or pipes, and various phenomena of RBCs behavior, such as margination and segregation,\cite{Chen2019,Kumar2012} self-organization in confined flows,~\cite{Aouane2017, Iss2019a, Lazaro2019} and cell focusing and separation\cite{Nieuwstadt2011, Mach2010, Tripathi2015} have extensively been studied. 

However, blood vessels in the circulation have more complex geometries and cross-sections (\textit{e.g.} sudden constrictions and expansions in stenosed arteries) that lead to cell depleted zones,\cite{Zhao2008} changes in cell density distribution,\cite{Vahidkhah2016} and result in clustering of microparticles in constricted blood flow.\cite{Bacher2017} To mimic pathophysiological geometric blockages and obstructed blood vessels that are associated with blood vessel diseases, constricted microfluidic channels with abrupt contractions and expansions are often used. Faivre~\etal\cite{Faivre2006} experimentally probed the flow of RBCs through microfluidic constrictions and showed that rapid variations of the channel cross-section dramatically change the lateral distribution of RBCs along the channel width. Additionally, constrictions can also lead to an increase in apparent blood viscosity\cite{Vahidkhah2016} and alter the thickness of the cell-free layer.\cite{Fujiwara2009} However, knowledge about the three-dimensional, spatial distribution of cells in the cross-section of constricted microfluidic devices remains vague.

In biomedical microfluidic applications, constricted channels are commonly used for cell and plasma separation.\cite{Faivre2006,Sollier2010,Lee2011a,Lee2011b} Generally, migration and focusing of cells or particles in straight microfluidic channels is induced by inertial lift forces\cite{Amini2014} or can be achieved by deformability based cell margination (\textit{e.g.} of rigid malaria-infected RBCs).\cite{Hou2010} In constricted microfluidic channels, particle focusing can be induced by secondary flows, generated by the sudden expansion. Park~\etal\cite{Park2009,Park2009a} used this effect to achieve continuous particle focusing in a multi-orifice contraction-expansion flow cell. The authors were able to control size-based particle separation at specific lateral positions across the channel width. However, they did not resolve the particle distribution across the channel depth. Further, Sollier~\etal\cite{Sollier2010} demonstrated that such constricted microchannels can provide a fast way for continuous plasma extraction from whole human blood. They showed that the cell-free layer increases post-contraction, induced by microvortices downstream of the expansion. Wang\cite{Wang2013} numerically investigated particle trapping and focusing in a constricted microfluidic device. For rectangular channels with high aspect ratios $W/H>3$, he predicted strong particle focusing near the long faces of the top and bottom channel walls and near the shorter side walls, while the particle concentration strongly decreases towards the channel center. Nevertheless, focusing effects of cells in contraction-expansion geometries are poorly investigated. Particularly, knowledge about the three-dimensional distribution of RBCs in the channel cross-section, resolved along the channel width and depth, in non-confined flows is still missing. Further, the effect of cell deformability on focusing remains poorly understood, which can help to improve high-throughput microfluidic cell sorting and plasma extraction.

In this study, we therefore want to answer the following specific questions; (a) How is the three-dimensional distribution of RBCs influenced when passing through a sudden constriction?, (b) Do deformable and non-deformable (\textit{e.g.} healthy and non-healthy) RBCs exhibit different flow behavior in contraction-expansion channels?, and (c) What is the effect of a non-uniform cell distribution on density-weighted averaging velocimetry techniques (\textit{e.g.} standard particle image velocimetry)?

We investigate living and rigid RBC suspensions in a constricted microfluidic device with a large cross-section ($400\times 52~\mu$m$^2$). Individual cells are first tracked in multiple layers across the channel depth. We then resolve the spatial distribution of RBCs in the channel cross-section and demonstrate that it dramatically changes when flowing through the abrupt constriction. We discuss how such a non-uniform distribution of cells results in an apparent double-peaked velocity profile post-contraction when averaging over the channel height. Additionally, we show that rigid RBCs linger in the microvortices downstream of the constriction, which can be used for cell classification.\cite{Hur2011, Chen2019}

%%%%%%%%%%%%%%%%%%%%%%%%%%%%%%%%%%%%%%%%%%%%%%%%%%%%%%%%%%%%%%%%%%%%%%%%%%%%%%%%%%%%%%%%%%%%%%%%%%%%%%%%%%%%%%
\section{Materials and methods}
\subsection{RBCs solution preparation}
Capillary blood is taken with informed consent from healthy voluntary donors into EDTA coated tubes and resuspended in phosphate buffered saline solution (PBS, Gibco). RBCs are extracted after centrifugation at $1,500~g$ for 5~minutes from the sediment and are washed with PBS. This procedure is repeated three times and final hematocrit concentrations are adjusted to $1\%~Ht$ and $5\%~Ht$. In order to detect individual cells in particle tracking measurements and resolve the distribution of RBCs in the channel cross-section, we limit the investigations in this study to concentrations below $5\%~Ht$, much lower than physiologically relevant in the circulation.\cite{Secomb2017} Above $5\%~Ht$, RBCs cannot be discriminated automatically, due to crowding.\cite{Iss2019a} Hence, the RBC suspensions examined here, have concentrations commonly studied in microfluidic experiments.

Further, we investigate the effect of cell deformability on the flow behavior of red blood cells in constricted channels. Therefore, RBCs are rigidified using $0.1\%$ glutaraldehyde (Sigma-Aldrich).\cite{Abay2019} Solutions are incubated for one hour and subsequently washed three times with PBS to remove remaining glutaraldehyde. Stiffened RBCs suspensions are prepared at $1\%~Ht$ and $5\%~Ht$. Blood withdrawal and preparation as well as all experiments are performed according to regulations and protocols that were approved by the ethic commission of the `Aerztekammer des Saarlandes' (reference No 24/12). Study participants were fully informed regarding the purposes of the study and consent was obtained.

\subsection{Microfluidic setup}
A microfluidic device is fabricated using polydimethylsiloxan (PDMS) (Momentive Performance Materials) through standard soft lithography.\cite{Friend2010} Inlet and outlet channels are punched in the PDMS chip, which is then bonded to a glass slide using a plasma cleaner (Harrick Plasma).

\begin{figure}[h]
\centering
  \includegraphics[width=8.3cm]{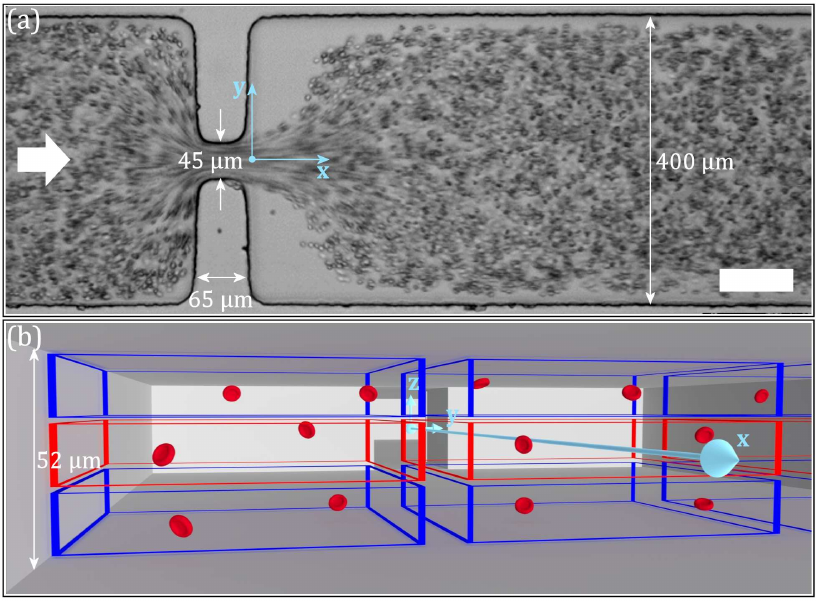}
  \caption{(a) Microscopic snapshot of a living RBC suspension ($5\%~Ht$) flowing through the microfluidic contraction-expansion channel at $\Delta p=140$~mbar. The flow direction is from left to right, as indicated by the white arrow. (b) Schematic representation of the regions of interest (ROI) in which RBCs are tracked in the channel middle and close to the side walls in $y$-direction, using a $60\times$ lens with $NA=1.25$. Blue boxes correspond to ROIs close to the top and bottom and red boxes indicate ROIs in the middle of the channel in $z$-direction. The origin of the coordinate system is in the center of the channel width at the contraction exit and in the middle of the channel height, see cyan arrows.}
  \label{FIG1}
\end{figure}

The microfluidic channel has a total length of $L=30$~mm, with a rectangular cross-section of width $W=400~\mu$m and height $H\approx52~\mu$m, similar to previous studies on particle focusing.~\cite{Park2009,Park2009a} The middle of the channel contains a sudden contraction-expansion with a width $W_{\text{c}}\approx45~\mu$m and a length $L_{\text{c}}\approx65~\mu$m. Figure~\ref{FIG1}(a) shows a microscopic snapshot of a RBC suspension ($5\%~Ht$) flowing through the channel center part. We define the origin of the coordinate system at the contraction exit in the channel middle in $y$-direction (cyan arrows in Fig.~\ref{FIG1}) and at the middle of the channel height $H/2$ in $z$-direction. The suspension enters the channel through the inlet, located at $x=-15$~mm upstream, passes through the contraction-expansion in positive $x$-direction, as indicated by the white arrow in Fig.~\ref{FIG1}(a), and exits the channel at $x=15$~mm downstream of the contraction.

The inlet and outlet are connected with micro medical grade polyethylene tubing (0.86 mm inner diameter (ID) and 1.32 mm outer diameter (OD), Scientific Commodities Inc.). In order to prevent cells from adhering to the inner surfaces, channel and tubing are coated with BSA (1~mg/mL bovine albumin, Sigma-Aldrich)  prior to the experiments. RBC suspensions are driven through the channel using a high-precision pressure device (Elveflow OB 1 MK 3) by applying a broad range of pressure drops $20\leq\Delta p\leq160$~mbar between the inlet and outlet.
 
The microfluidic device is mounted on an inverted microscope (Eclipse TE2000-S, Nikon) equipped with a high-speed camera (MEMRECAM GX1, NAC, up to 2559 frames per second at full resolution of $1280 \times 1024$ pixels) and LED illumination (ZLED CLS 9000 MV-R, Zett Optics). The frame rate of the high-speed camera is chosen such that cell displacement between two images is around 4 pixels. A $60\times$ oil-immersion objective (CFI Plan Fluor, Nikon) with high numerical aperture $NA=1.25$, as well as a $10\times$ air objective (Plan Fluor Nikon) with $NA=0.3$ are used to capture images of the RBCs flowing through the channel. The measurement depth $\delta z_{\text{m}}$ over which cells are detected and contribute to the determination of the velocity field is $\delta z_{\text{m}}\approx15~\mu$m for the $60\times$ lens and $\delta z_{\text{m}}\approx70~\mu$m for the $10\times$ lens.\cite{Meinhart2000} We use the $60\times$ objective ($\delta z_{\text{m}}/H\approx0.3$) to resolve the position of RBCs in three layers across the channel height $H$: close to the bottom and top ($z\approx\pm 15~\mu$m) and in the middle ($z=0$) of the channel, corresponding to the blue and red boxes in Fig.~\ref{FIG1}(b), respectively. Here, we focus on the border regions ($\pm200\leq y \leq \pm80~\mu$m) or on the center part ($-60\leq y \leq 60~\mu$m) of the channel width with respect to the lateral $y$-direction and with regions of interest (ROI) of roughly $120\times 120~\mu$m$^2$, which accounts for $0.3~W$. Furthermore, using the $10\times$ objective allows us to cover the complete channel width ($-200\leq y \leq 200~\mu$m) and to average the velocity of RBCs across the whole channel height ($\delta z_{\text{m}}/H\approx1.4$). All experiments are performed at room temperature. The tracking of RBCs as well as particle image velocimetry of the RBC suspension are performed at multiple $x$ positions upstream (pre-contraction) and downstream (post-contraction) of the contraction-expansion in the channel.

\subsection{Particle tracking velocimetry}
At RBC concentrations below $5\%~Ht$ we perform particle tracking velocimetry (PTV) by detecting the position of individual cells. Due to adjusting the exposure time to be much shorter than the frame rate, cells within the focus are imaged sharp and no motion blur occurs, while cells outside the focal plane are blurred. A self-written \textsc{Matlab} program is used to subtract the common background of all images and detect the positions of individual cells in each image. The spatial gradient of the images is calculated to distinguish between sharp and blurred cells. The resulting images in gradient magnitudes are binarized by a threshold value. Therefore, only objects in focus with sharp edges are detected. The positions of detected cells over the image sequence are linked and result in individual trajectories. From those trajectories, individual velocities as function of space and time are determined, taken the time derivative. The combination of all cell velocities results in the time-averaged velocity field. Further, the RBC distribution as number density is calculated by counting all sharp cells in the binned $y$-direction, dividing by the number of frames in the image sequence and the corresponding bin volume.

\subsection{Particle image velocimetry}
We use two-dimensional particle image velocimetry (PIV) as an additional method to analyze the flow in the microfluidic device. PIV is commonly used to determine the density-weighted average velocity of an particle ensemble in interrogation areas, in contrast to the determination of the velocities of individual particles using the PTV technique. In PIV measurements, images are captured in pairs, using a $10\times$ objective with $NA=0.3$. Here, RBCs are used as flow tracers, without the addition of further particles.\cite{Poelma2012, Passos2019} We use an open source PIV software\cite{Thielicke2014a} for quantitative analysis of the flow field and velocity vectors are obtained in interrogation areas of $32\times32$~pixels in $x$- and $y$-direction. With the volume illumination setup used here, the flow of RBCs over the entire channel height $H$ is recorded. Hence, data is substantially averaged over the velocity gradient in $z$-direction. The effect of this density-weighted averaging in PIV analysis, in combination with an non-uniform cell distribution in the channel cross-section, is discussed and compared with PTV measurements in section~3.3 and 3.4.

%%%%%%%%%%%%%%%%%%%%%%%%%%%%%%%%%%%%%%%%%%%%%%%%%%%%%%%%%%%%%%%%%%%%%%%%%%%%%%%%%%%%%%%%%%%%%%%%%%%%%%%%%%%%%%
\section{Results and discussion}
Representative images of RBC suspensions passing the constriction at different pressure drops are provided in Fig.~S1 (ESI$\dag$), similar to Fig.~\ref{FIG1}(a). In the corner regions downstream of the contraction, we observe cell-free zones with recirculating flows. The size of these vortices increases with increasing pressure drop. The formation of such microvortices in sudden expansions leads to secondary flows that have been studied before and that are commonly used for separation and fractionation.\cite{Sollier2010, Chiu2007, Park2009, Park2009a} 

Here, we discuss the influence of this effect on the flow and spatial distribution of RBCs in the channel cross-section. Representative results of living or rigid RBCs are shown in this section, while complementary data is provided in the ESI$\dag$. Using PTV, we first resolve the velocities and distribution of RBCs in different layers across the channel depth in section~3.1. Second, the spatial cell distribution derived from PTV across the whole channel width is discussed in section~3.2. In section~3.3 and section 3.4, we use PIV to study the apparent mean velocity profile as a function of pressure drop and along the channel length, respectively. Finally, the trapping of RBCs in the vortex regions downstream of the constriction is discussed in section~3.5.

\subsection{RBC velocities and distribution in different $z$-layers}
\begin{figure}[h]
\centering
  \includegraphics[width=8.3cm]{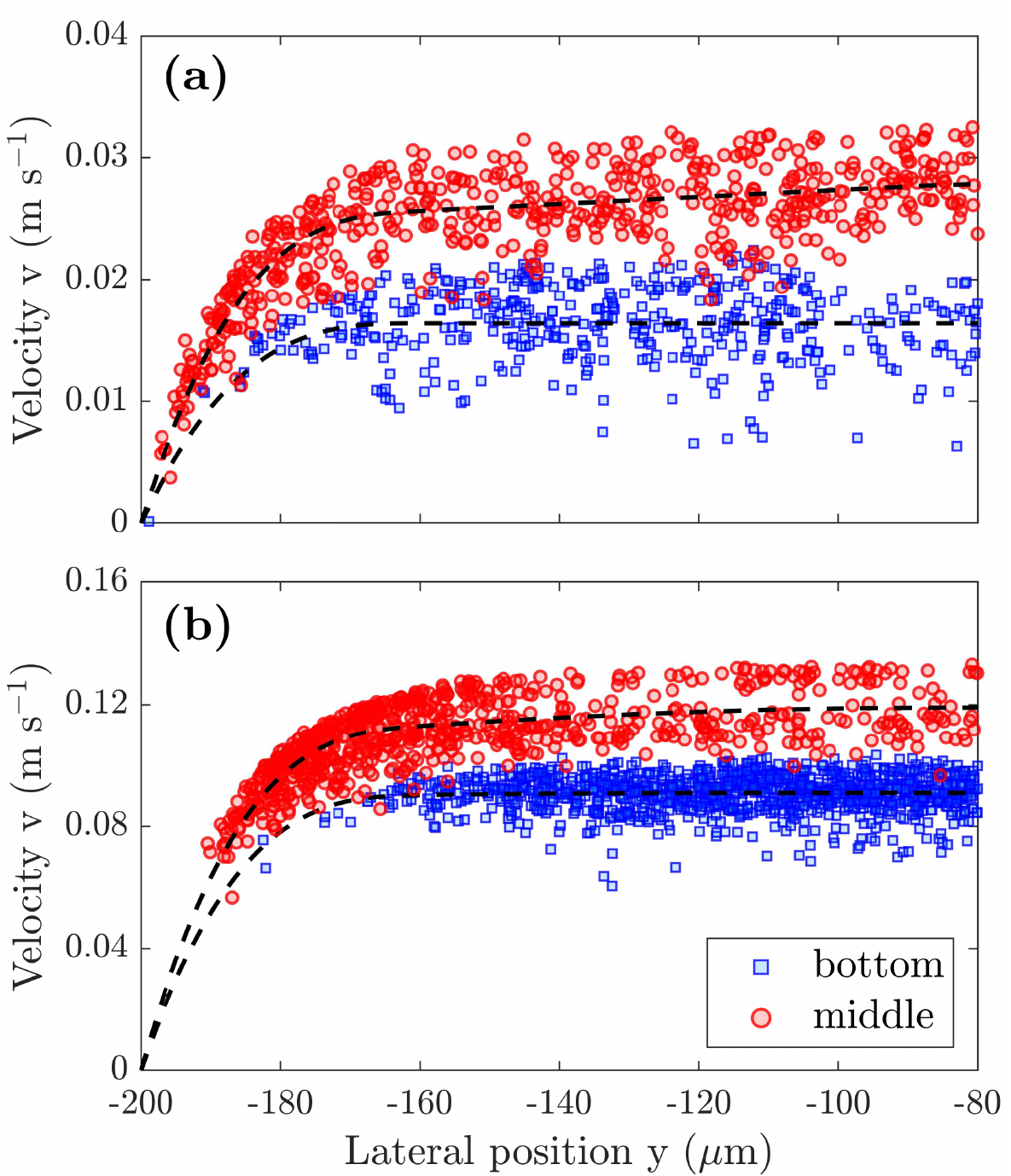}
  \caption{Velocities of rigid RBCs ($1\%~Ht$) $x=10$~mm post-contraction at a pressure drop of (a) $\Delta p=40$~mbar and (b) $\Delta p=140$~mbar. Cell positions are detected in two layers, close to the bottom ($z\approx-15~\mu$m) and in the channel middle ($z=0$), represented as blue and red symbols, respectively. Dashed lines represent the mean velocities at each plane, calculated using the SLM (Shape Language Modeling) tool implemented in \textsc{Matlab} with the boundary condition that the velocity is zero at the channel wall $y=-200~\mu$m.}
  \label{FIG2}
\end{figure}

Figure~\ref{FIG2} shows the velocities of individual rigid RBCs ($1\% ~Ht$) at a representative (a) low and (b) high pressure drop. Data is acquired close to the channel border in $y$-direction ($-200\leq y \leq -80~\mu$m) in the bottom and middle layers of the channel, $x=10$~mm post-contraction. Note that the channel border corresponds to the side walls of the channel at $y=-200~\mu$m.

At both pressure drops, RBCs flow in a plug-like velocity profile along the $y$-direction, due to the high aspect ratio of the channel $W/H\approx8$. The average velocity in the bottom or middle plane is constant throughout most of the channel width ($-160\leq y \leq 160~\mu$m). The corresponding velocities in the center part of the channel ($-60\leq y \leq 60~\mu$m) are shown in Fig.~S2, ESI$\dag$. The scattering of the velocity data results from data acquisition over $\delta z_{\text{m}}\approx15~\mu$m, hence over the parabolic velocity gradient in $z$-direction.

Besides the increase of velocity magnitude with increasing pressure, the main difference between $\Delta p=40$~mbar and $\Delta p=140$~mbar is the depletion of RBCs close to the channel wall ($y=-200~\mu$m). While cells flow in close proximity of the channel wall in the middle of the channel at $\Delta p=40$~mbar (red symbols in Fig.~\ref{FIG2}), a pronounced cell-free layer forms at $\Delta p=140$~mbar adjacent to the channel wall ($-200\leq y \leq -190~\mu$m). The magnitude of this cell-depleted zone increases with increasing pressure drop, as seen in Fig.~S3 (ESI$\dag$). Here, the cell-free layer is larger for living RBCs than for rigid RBCs.\cite{Fujiwara2009} Note that the cell-depleted layer upstream of the contraction seems to be independent of the applied pressure drop and is less pronounced in y-direction than downstream (see Fig.~\ref{FIG1}(a) and Fig.~S1 in ESI$\dag$), in agreement with previous investigations.\cite{Pinho2013, Faivre2006, Sollier2010, Fujiwara2009} In the bottom plane (blue symbols in Fig.~\ref{FIG2}), the strong interactions with the channel side and bottom wall in the corner regions force the cells to migrate away even further than in the channel middle, resulting in an increase of the cell-free zone in the bottom layer ($-200\leq y \leq -170~\mu$m in Fig.~\ref{FIG2}(b)).

The formation of the cell-depleted layer near the walls is the main reason for the distinctive flow properties of blood in microvessels.\cite{Secomb2017} While this phenomenon has extensively been studied by averaging over the total channel height,\cite{Faivre2006, Kim2009, Sollier2010, Fujiwara2009, Pinho2013} we resolve the spatial distribution and velocities of RBCs across the channel depth in this study. The strong hydrodynamic interactions between cells and the walls result in a more pronounced cell-depleted zone in the vicinity of the channel corners.\cite{Chen2019, Amini2014, DiCarlo2009}

\begin{figure}[h]
\centering
  \includegraphics[width=8.3cm]{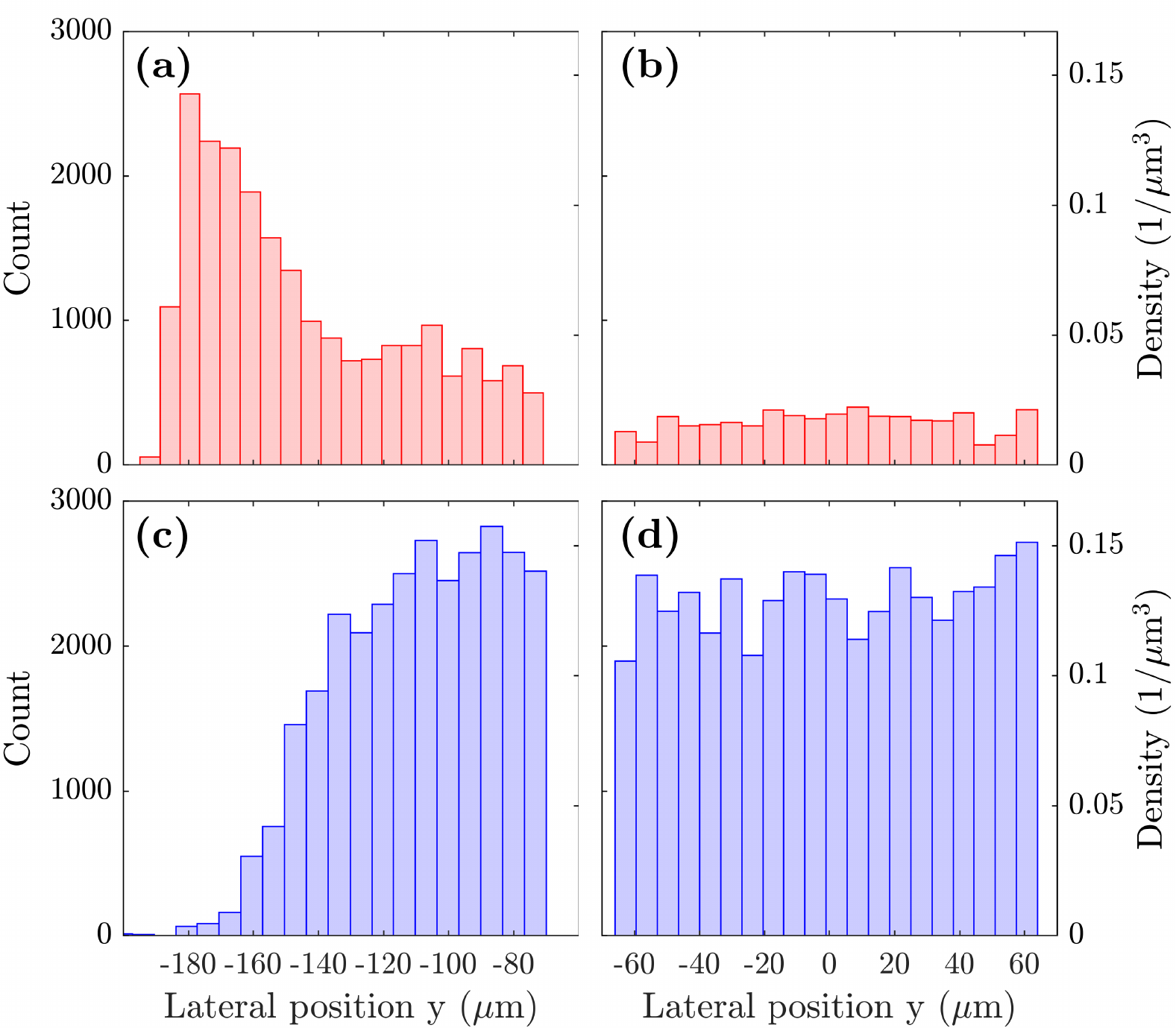}
  \caption{Distribution of rigid RBCs ($1\%~Ht$) $x=10$~mm post-contraction along the channel width in the channel middle (upper row), and close to the bottom (lower row) in $z$-direction. (a) and (c) represent the channel border, i.e. close to the channel side wall at $y=-200~\mu$m, and (b) and (d) show the channel center in $y$-direction. The pressure drop is $\Delta p=140$~mbar. Data is accumulated over 4,000 frames.}
  \label{FIG3}
\end{figure}

In an effort to emphasize the distribution of RBCs in y-direction downstream of the contraction, we plot the accumulated number of cells as a function of the channel width in Fig.~\ref{FIG3} for $\Delta p=140$~mbar. Here, the top row corresponds to the channel middle, while the lower row corresponds to the channel bottom. Further, the panels (a) and (c) show the distribution close to the channel border ($-200\leq y \leq -80~\mu$m) near the channel side wall at $y=-200~\mu$m, while (b) and (d) correspond to the channel center ($-60\leq y \leq 60~\mu$m). The right $y$-axis of Fig.~\ref{FIG3} shows the number density of RBCs ($1/\mu$m$^3$), passing the ROI at a certain position along the channel width over data acquisition.

In Fig.~\ref{FIG3}(c) the depletion of RBCs in the channel bottom at the corner region $-200\leq y \leq -160~\mu$m is clearly visible, similar to Fig.~\ref{FIG2}(b). In the rest of the channel ($-120\leq y \leq 60~\mu$m), the distribution of cells is uniform across the width in $y$-direction. On the other hand, a dramatic difference arises in the channel middle. Here, a strong accumulation of RBCs is observed close to the border ($-190\leq y \leq -150~\mu$m) at the short channel side face, as seen in Fig.~\ref{FIG3}(a). In contrast to this strong focusing, the number of RBCs drastically decreases towards the channel center ($-60\leq y \leq 60~\mu$m), where considerably fewer cells pass the middle plane, as seen in Fig.~\ref{FIG3}(b). Further, the amount of RBCs in the channel middle (Fig.~\ref{FIG3}(b)) is significantly reduced compared to the bottom layer (Fig.~\ref{FIG3}(d)) at the same lateral $y$-position. Note that the top layer exhibits a similar cell distribution as the bottom layer, as shown in Fig.~S4, ESI$\dag$. At low pressure drops, this focusing phenomenon of RBCs post-contraction is hardly detectable close to the experimental resolution limit, as exemplified for $\Delta p=40$~mbar in Fig.~S5, ESI$\dag$. Furthermore, we observe a uniform distribution of RBCs across the channel width upstream of the constriction, as shown in Fig.~S6 (ESI$\dag$), where the RBC distribution is plotted at $x=10$~mm pre-contraction, complementary to the data at $x=-10$~mm post-contraction in Fig.~\ref{FIG3}.

Figure~\ref{FIG3} demonstrates that RBCs preferentially flow near the four faces of the channel walls post-contraction, \textit{i.e.} the wider faces at the bottom and top and the short faces at the side walls. In the constricted microfluidic channel used here, RBC focusing is induced by secondary flows downstream of the abrupt expansion.\cite{Park2009, Park2009a} However, migration and focusing of cells or particles can also be achieved in straight microfluidic channels, induced by inertial lift forces.\cite{Amini2014} At a finite Reynolds number $\text{Re}$, which relates the inertial to viscous forces, two lift forces act on a particle in a Poiseuille flow: the wall induced lift force, due to particle-wall interactions that pushes the particle away from the wall, and the shear-gradient induced force, due to the curved velocity profile that leads to a movement away from the channel center. Based on the balance between these two opposing forces, different equilibrium particle positions in the channel cross-section can be achieved.\cite{DiCarlo2009}

Lateral migration and a focusing of particles, induced by inertial lift forces~\cite{Segre1961} occurs in straight rectangular channels for particle Reynolds numbers $\text{Re}_{\text{p}}$ than unity.\cite{Amini2014, DiCarlo2009} The particle Reynolds number is defined as 

\begin{equation}
  \text{Re}_{\text{p}}=\text{Re}_{\text{c}} \frac{a^2}{D_\text{h}^2}=\frac{\rho \text{v}_{\text{m}} a^2}{\eta D_{\text{h}}}
\end{equation}

\noindent where $\text{Re}_{\text{c}}$ is the channel Reynolds number, $\rho$, $\text{v}_{\text{m}}$, $a$, $\eta$, and $D_{\text{h}}$ are the fluid density, the maximum flow velocity, the particle or cell diameter, the fluid viscosity and the hydraulic diameter of the microfluidic channel, respectively. The hydraulic diameter for a rectangular channel is defined as $D_{\text{h}}=2WH/(W+H)$, where we use the width $W=400~\mu$m to calculate $\text{Re}_{\text{p}}$. However, note that significantly higher $\text{Re}_{\text{c}}$ and $\text{Re}_{\text{p}}$ can be achieved in the constricted part, where the small width of $W_{\text{c}}\approx45~\mu$m results in much higher velocities.

Generally, two stable equilibrium positions near the wider channel faces exist in rectangular channels, without the four-fold symmetry of a square, due to the stronger shear-gradient lift forces in $z$-direction. Thus, particles or cells are pushed away from the channel centerline toward the wider sides. Here, the balance between shear-gradient and wall lift forces results in stable equilibrium positions along the wider faces. This effect is much less pronounced towards the shorter faces, since the plug-like velocity profile leads to much weaker shear gradient lift forces in $y$-direction.\cite{Amini2014} However, increasing $\text{Re}_{\text{p}}$ results in a total number of four focusing positions along both the wide and short faces.\cite{Hur2010,Ciftlik2013} Hur~\etal\cite{Hur2010} investigated the ordering of beads and blood cells in rectangular channel ($W=16~\mu$m and $H=37~\mu$m) and observed that particles accumulate close to all four faces at $\text{Re}_{\text{p}}\approx4.5$. Much higher values of $\text{Re}_{\text{p}}\approx150$ were reported by Ciftlik~\etal\cite{Ciftlik2013} for particle focusing of $10~\mu$m beads at all four channel faces in a rectangular channel with $W=50~\mu$m and $H=80~\mu$m. For the data shown in Fig.~\ref{FIG3}, the Reynolds numbers are $\text{Re}_{\text{p}}\approx0.1$ and $\text{Re}_{\text{c}}\approx12$, with $\rho=1$~g/cm$^3$, $a=8~\mu$m, and $\eta=1$~mPa~s. This $\text{Re}_{\text{p}}$ value is much smaller than those reported by Hur~\etal\cite{Hur2010} and Ciftlik~\etal\cite{Ciftlik2013}, and also smaller than unity, which indicates that cell focusing solely due to inertial lift forces would not occur in the straight channel pre-contraction. Further, a certain length $L_{\text{f}}=\pi \eta D_{\text{h}}^2/(\rho U_{\text{m}} a^2 f_{\text{L}})$, with the non-dimensional lift coefficient $f_{\text{L}}=0.05$, along the flow direction is required for particles to reach stable equilibrium positions inside rectangular channels.\cite{Amini2014} In order to reach equilibrium positions in the experiments shown here, a length of $L_{\text{f}}\approx60$~mm would be required, hence much longer than the entry length ($15$~mm) upstream of the contraction. Thus, we do not expect any inertial focusing to arise pre-contraction, in agreement with our experimental results, shown in Fig.~S6. The observed focusing phenomenon shown in Fig.~\ref{FIG3} emerges as a consequence of the constricted flow geometry, which is discussed below.

\subsection{Change of RBC distribution along the flow direction}
\begin{figure}[h]
\centering
  \includegraphics[width=8.3cm]{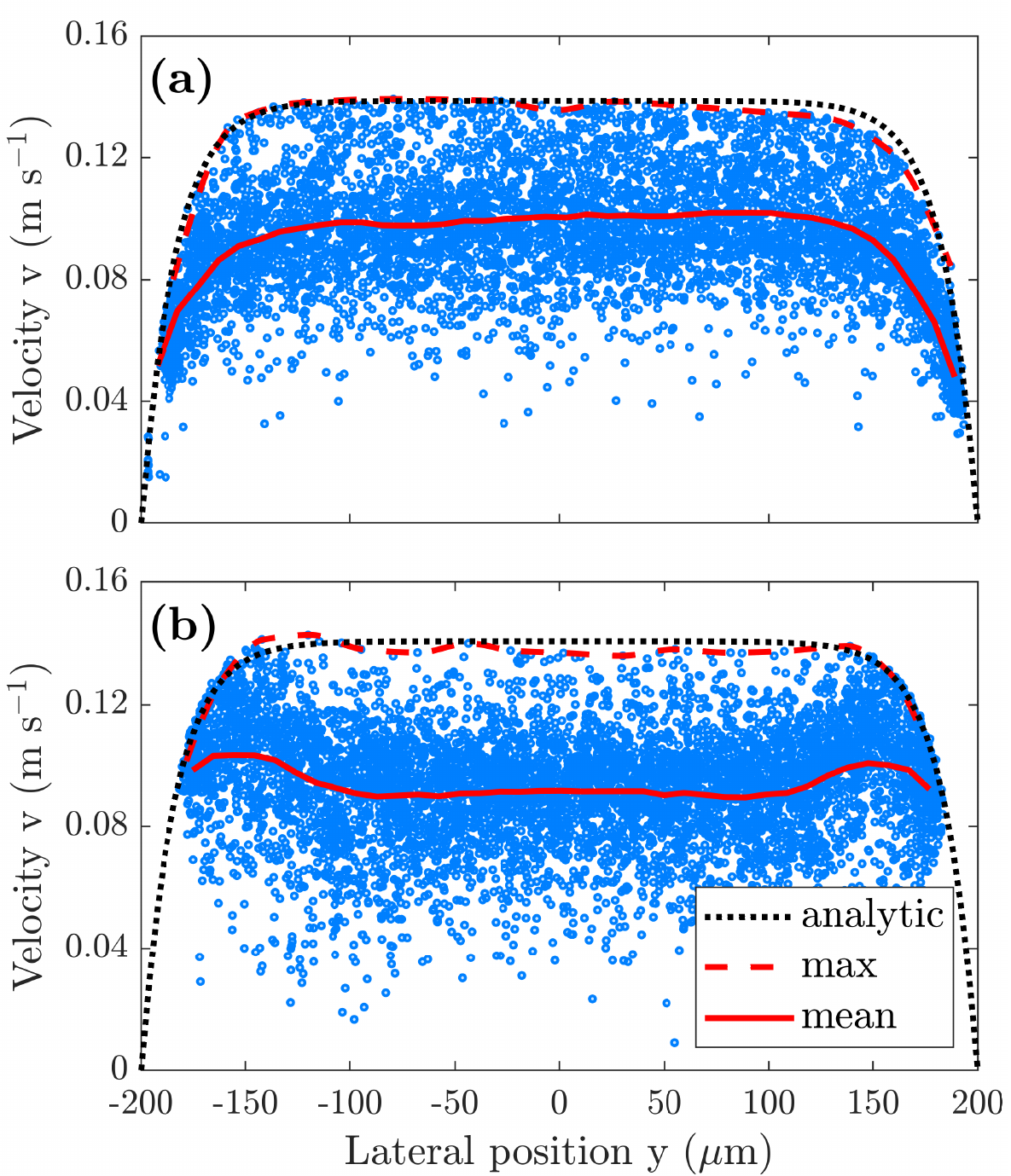}
  \caption{PTV: Velocities of individual living RBCs ($1\%~Ht$), plotted as blue dots, at $\Delta p=140$~mbar (a) $x=-10$~mm pre-contraction and (b) $x=10$~mm post-contraction. Data is acquired across the whole channel height, using a $10\times$ lens with $NA=0.3$. The dashed and solid red lines represent the maximum and mean velocities of all cells along the channel width, respectively. The dotted black line indicates the velocity profile for a Newtonian fluid across the channel middle $z=0$. }
  \label{FIG4}
 \end{figure}

In the previous section, we revealed the distinctive spatial distribution of RBCs in different layers across the channel depth, focusing on small ROIs in the lateral y-direction. Here, we discuss the change of cell distribution induced by the constriction. Using a lower magnification $10\times$ lens with $NA=0.3$ allows us to track RBCs across the whole channel width $W$, as well as to detect all cells throughout the channel height $H$. Figure~\ref{FIG4} shows the velocities of individual living RBCs that pass the microfluidic channel (a) $x=-10$~mm pre-contraction and (b) $x=10$~mm post-contraction. Additionally, we plot the maximum and mean velocities at a certain lateral $y$-position across the channel width as red dashed and solid lines, respectively. The mean velocity at a specific $y$-position is calculated by density-weighted averaging the velocities of RBCs, flowing at different layers across the channel height $H$, at this $y$-position. The dotted black line indicates the velocity profile across the channel middle $z=0$ for a Newtonian fluid, with the same viscosity and flowing through a rectangular channel under the same experimental conditions.~\cite{Blythman2017} The corresponding data for rigid cells is provided in Fig.~S7 (ESI$\dag$).

Upstream of the contraction, RBCs flow in a plug-like velocity profile across the channel width, as shown in Fig.~\ref{FIG4}(a). Detecting cells across the velocity gradient in $z$-direction yields a broad velocity distribution at a fixed lateral y-position. Here, slower cells correspond to planes closer to the top and bottom of the channel, while the fastest cells flow in the channel middle, as shown in Fig.~\ref{FIG2}. The enveloping line of the fastest particles corresponds to the maximum velocity (dashed red line) at $z=0$ and is in good agreement with the analytic solution for the velocity profile in the channel middle of a Newtonian fluid (dotted black line). Further, the plug-like shape of the mean velocity (solid red line) indicates that the distribution of RBCs is uniform along the channel depth $H$. This homogeneous RBC distribution pre-constriction is also found in particle tracking experiments, resolved along the channel height, as previously discussed (see Fig.~S6 (ESI$\dag$)).

However, passing through the sudden constriction, a significant change of the cell distribution is observed, as shown in Fig.~\ref{FIG4}(b). Fewer cells pass the center part $-100\leq y \leq 100~\mu$m in the channel middle, where the velocity is largest. In contrast, we detect an accumulation of RBCs close to the side walls at $y\approx\pm160~\mu$m. In these regions, the majority of cells are located near the channel middle plane $z=0$. This effect is also observed in Fig.~\ref{FIG2}(b) and leads to a double-peaked mean velocity profile, when averaging over the channel depth (solid red line in Fig.~\ref{FIG4}(b)). It is important to note that this apparent double-peaked profile arises as a consequence of density-weighted averaging in combination with a non-uniform distribution of flow tracer, \textit{i.e.} RBCs. The maximum velocity across the channel width still matches the analytic solution for a Newtonian fluid. However, since considerably fewer RBCs flow in the channel middle at $z=0$, compared to the channel bottom and top at $z=\pm15~\mu$m (see Fig.~\ref{FIG2}(b)), averaging over the channel depth results in a lower mean velocity in the channel center $-100\leq y \leq 100~\mu$m. For rigid RBCs, we observe the same flow behavior and also find an apparent double-peaked mean velocity profile, as shown in Fig.~S7 (ESI$\dag$).

Recently, Iss~\etal\cite{Iss2019a} investigated the self-organization of RBCs under confined flows and reported a non-uniform cross-flow distribution of cells. The authors also observed the formation of bands with high local RBC concentration at specific lateral $y$-position, depending on $\Delta p$. However, in this case, the non-homogeneous RBC distribution emerged as a consequence of confinement in the flat channel ($H\approx9~\mu$m), in contrast to the constricted deeper channel ($H\approx52~\mu$m) used in this study.

Another phenomenon that can be seen in Fig.~\ref{FIG4} is the increase of the cell-depleted layer, induced by the constriction. While cells pass in close proximity of the channel side walls ($y=\pm200~\mu$m) pre-contraction, the cell-free layer increases downstream of the contraction, as reported before.\cite{Pinho2013, Faivre2006, Sollier2010, Fujiwara2009} Here, we find that the cell-depleted layer spans laterally up to $20~\mu$m at each side of the channel in y-direction.

The most commonly used technique to visualize flow in microfluidic devices is PIV. This method allows us to obtain quantitative velocity data, based on the displacement of an ensemble of flow tracers within interrogation areas.\cite{Raffel2018} Using volume illumination in combination with a low $NA$ lens, the depth of field can become larger than the channel depth and thus results in a potential averaging of velocities along the gradient in $z$-direction,\cite{Meinhart2000} similar to the mean velocity values (solid red lines in Fig.~\ref{FIG4}) obtained from PTV. In the next section, we employ PIV analysis on the flow of RBCs and demonstrate how the non-uniform distribution of cells in the channel cross-section also yields a double-peaked velocity profile, similar to the particle tracking results shown in Fig.~\ref{FIG4}(b). 

\subsection{RBC focusing with increasing pressure drop}
\begin{figure}[h]
\centering
  \includegraphics[width=8.3cm]{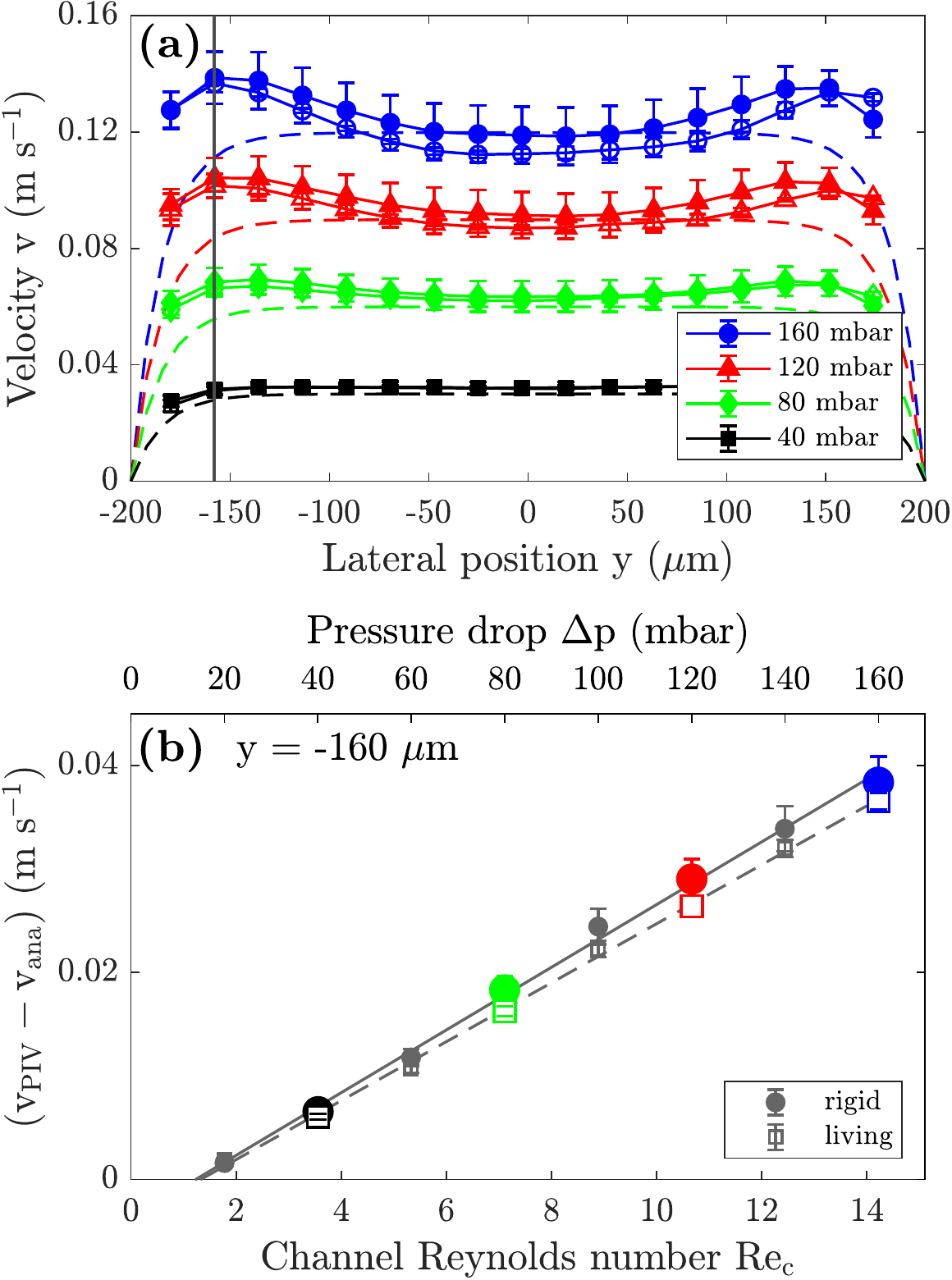}
  \caption{PIV: (a) Mean velocity profiles across the channel width $x=10$~mm post-contraction of rigid (solid symbols) and living (open symbols) RBCs ($5\%~Ht$) at different pressure drops, using a $10\times$ lens with $NA=0.3$. Dashed lines correspond to the analytical solution of the mean velocity profiles of a Newtonian fluid. With increasing pressure, the particle and channel Reynolds numbers are $\text{Re}_{\text{p}}=0.029$, $\text{Re}_{\text{p}}=0.058$, $\text{Re}_{\text{p}}=0.087$, and $\text{Re}_{\text{p}}=0.12$, and $\text{Re}_{\text{c}}=3.6$, $\text{Re}_{\text{c}}=7.1$, $\text{Re}_{\text{c}}=11$, and $\text{Re}_{\text{c}}=14$, respectively. Error bars correspond to standard deviations of different measurements. (b) The difference between the velocity peak at $y=-160~\mu$m and the analytical solution of the velocity profile of a Newtonian fluid as a function of pressure drop and channel Reynolds number. Colored symbols correspond to data from (a). The solid and dashed gray lines correspond to linear fits of the rigid and living RBC data, respectively.}
  \label{FIG5}
\end{figure}

Figure~\ref{FIG5}(a) shows the average velocity profiles $x=10$~mm downstream of the contraction for rigid (closed symbols) and living (open symbols) RBCs ($5\%~Ht$), determined through PIV analysis. The dashed lines correspond to theoretical predictions of the mean velocity profiles, which are derived from an analytical solution of the flow profile for a Newtonian fluid, averaged across the channel depth.\cite{Blythman2017} Figure~\ref{FIG5}(b) shows the velocity peak height, \textit{i.e.} the velocity from the PIV measurements $\text{v}_{\text{PIV}}$ at $y=-160~\mu$m minus the analytical mean velocity $\text{v}_{\text{ana}}$ at the same $y$-position, as indicated by the vertical black line in Fig.~\ref{FIG5}(a).

Both, living and rigid RBC suspensions follow a plug-like velocity profile at small pressure drop, \textit{e.g.} $\Delta p=40$~mbar in Fig.~\ref{FIG5}(a). Here, we do not observe any quantitative difference in the velocity profiles of living and rigid cells, hence both profiles overlap at small pressure drops. Further, at low pressure drops, the velocity profiles obtained through PIV analysis are in good agreement with the analytical predictions, as indicated by the dashed lines. However, we observe strong deviations from the theoretical mean velocity profiles with increasing pressure drop, resulting in a transition from a plug-like towards a double-peaked velocity profile. At $\Delta p=80$~mbar, two velocity peaks become visible at $y\approx\pm160~\mu$m, while the velocity in the channel center $-100\leq y \leq 100~\mu$m is still constant. The magnitude of the peaks increases with further increase of pressure drop and the average velocity profile obtained from PIV analysis of the RBC ensemble approximates the mean velocity profile derived from tracking of single RBCs, shown in Fig.~\ref{FIG4}(b). Figure~\ref{FIG5}(b) shows the difference of the velocity at the peak position and the value from the analytical solution for the velocity profile of a Newtonian fluid. While there is a good agreement at lower pressure drops, the difference increases linearly with the pressure drop due to the emergence of the apparent velocity peak that is a consequence of the increase of cell density at this position. The graph indicates that flow focusing starts at a critical channel Reynolds number $\text{Re}_{\text{c}}\approx1.3$. However, this value must be taken with care because this method does not measure the density difference directly. Nevertheless, our data strongly indicates that cell separation in the geometry is not only driven by inertia but that there is also a critical transition threshold.  

It is important to emphasize that the double-peak velocity profile emerges as a consequence of density-weighted averaging over the channel depth by employing PIV analysis, where different layers contain varying numbers of cells, as exemplified in Fig.~\ref{FIG2}(b). Thus, the depletion of cells in the channel center ($z=0$ and $-100\leq y \leq 100~\mu$m) in combination with a higher amount of cells close to the top and bottom ($z\approx\pm 15~\mu$m and $-100\leq y \leq 100~\mu$m), yields a lower apparent velocity in the channel center. On the other hand, the strong focusing of RBCs in the middle plane, close to the side walls ($z=0$ and $y\approx\pm 160~\mu$m) in combination with the cell-depleted layer in the top and bottom plane ($z\approx\pm 15~\mu$m and $\pm200\leq y \leq \pm160~\mu$m), as shown in Fig.~\ref{FIG2}, leads to overestimating of the density-weighted average velocity. We use these distinctive profiles of apparent velocity as an qualitative measure for RBC focusing in the microfluidic channel. 

Particle focusing in microfluidic contraction-expansion devices was reported before. Park~\etal\cite{Park2009, Park2009a} examined the distribution of microspheres with different sizes ($2\leq a \leq 15~\mu$m) flowing through a multi-orifice channel. They observed a focusing of beads at specific lateral positions starting at $0.01\leq Re_{\text{p}}\leq 0.63$, depending on particle size. The corresponding channel Reynolds number $\text{Re}_{\text{c}}\approx 7$ for the onset of focusing is in good agreement with the results shown in Fig.~\ref{FIG5}. The authors attributed the non-homogeneous particle distribution post-contraction to a combination of inertial lift forces and secondary flows, induced by vortex flows in the suddenly expanding channel. However, they did not resolve the focusing effect across the channel height.

Similarly, Wang\cite{Wang2013} numerically studied particle sorting in constricted microfluidic channels, covering a broad range of $0.02\leq \text{Re}_{\text{p}}\leq 20$, depending on particle size, and $8\leq \text{Re}_{\text{c}}\leq 160$. He reported a strong focusing at two narrow lines in the channels center plane ($z=0$). Additionally, particles concentrated near the wider faces at the bottom and top for aspect ratios $W/H > 3$, depending on the axial position in the microfluidic device. Here, our experiments confirm the spatial distribution in the channel cross-section, as predicted numerically by Wang.\cite{Wang2013}

Furthermore, we do not observe a significant difference in the velocity data and focusing of rigid and living RBCs in Fig.~\ref{FIG5}. Chen~\etal\cite{Chen2019} recently reported that migration of RBCs depends on cell deformability. They used a mixture of living and stiffed RBCs and found that the impaired deformability of RBCs decreases the wall lift force and the velocity force, thus resulting in a different equilibrium positions between living and rigid RBCs. In contrast to the experiments shown in this study, they used a longer ($L=30$~mm) straight microfluidic channel with a smaller rectangular cross-section ($100 \times 10~\mu$m$^2$). The authors only observed a focusing of rigid RBCs when they were suspended in a living RBC suspension, but not when suspended solely in PBS. Hence, margination leads to different equilibrium positions in the study by Chen~\etal\cite{Chen2019} Although, we do not observe significant differences in the velocity data of rigid and living RBCs post-contraction, dramatic differences between both cells arise in the vortex regions, as discussed in section~3.5.

In this study, the non-uniform cell distribution and depletion of cells in the channel center post-contraction, and hence the double-peaked velocity profiles derived from PTV (Fig.~\ref{FIG4}(b)) and from PIV (Fig.~\ref{FIG5}(a)), arise as a consequence of geometry-induced focusing and not as a result of RBCs sedimentation. Significant sinking of cells across the channel height, while traveling through the channel in $x$-direction, is clearly ruled out showing that the cell distribution in the top layer is similar to the bottom layer (Fig.~S4). This holds even for lower pressure drops (Fig.~S5), where sedimentation is more likely due to the lower flow rate. Furthermore, we perform RBC tracking and PIV experiments in a density matched solution, using OptiPrep\texttrademark~(iodixanol aqueous solution, Sigma-Aldrich), to avoid potential sedimentation and RBC sinking. Therefore, RBC suspensions are prepared using 35\% OptiPrep\texttrademark~and 65\% PBS to adjust the density of the solution to $\rho\approx1.11~$g/ml, matching RBC density.\cite{Losserand2019} Using density matched suspensions does not influence the cross-sectional focusing phenomenon and we also observe the emergence of apparent double-peaked velocity profiles in these suspensions in PIV analysis, as exemplified in Fig.~S8 (ESI$\dag$). Hence, we do not observe sinking and sedimentation of RBCs in the flow channel during the experiments.

\subsection{Evolution of the velocity profile}
Thus far, we examined the flow of RBC suspensions, mainly focusing on positions $x=\pm 10$~mm of the constriction, where the flow is quasi-stationary. Further, we perform PIV measurements at different $x$-positions along the channel flow direction, in an effort to probe the evolution of the mean velocity profile post-contraction. 

\begin{figure}[h]
\centering
  \includegraphics[width=8.3cm]{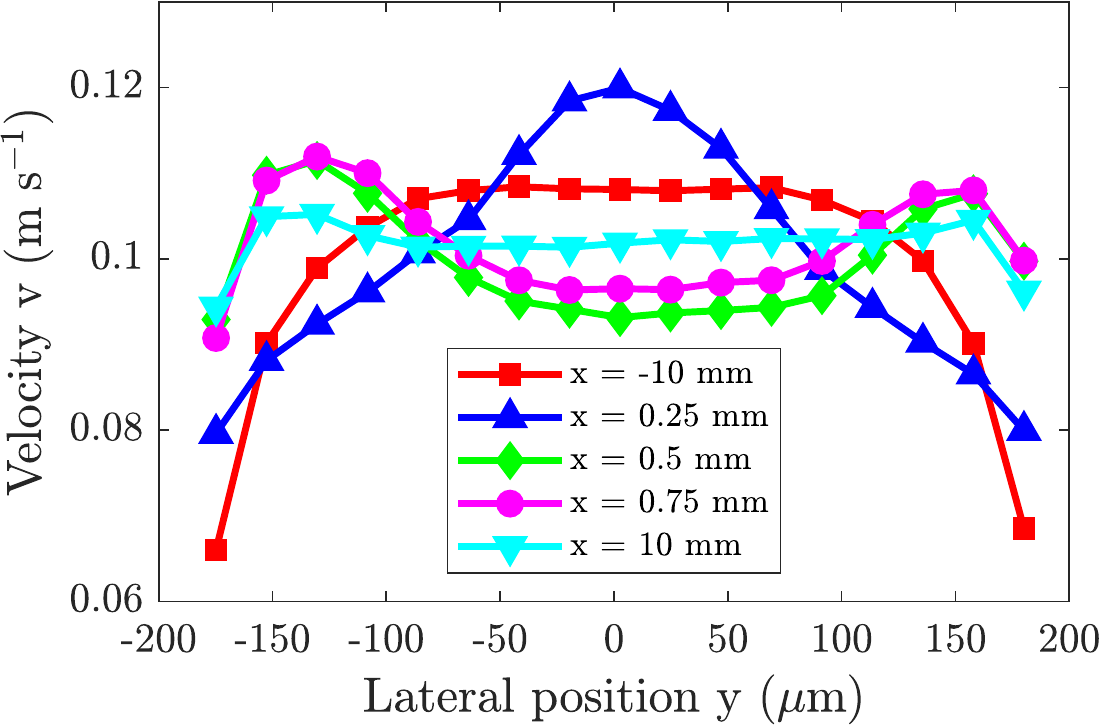}
  \caption{PIV: Mean velocity profiles across the channel width of a living RBC suspension ($5\%~Ht$) at different $x$ positions along the channel and at $\Delta p=120$~mbar.}
  \label{FIG6}
\end{figure}

Figure~\ref{FIG6} shows the mean velocity profiles of a $5\%~Ht$ suspension of living RBCs at different $x$-positions pre- and post-contraction. Here, the pressure drop $\Delta p=120$~mbar is chosen such that the non-uniform cell distribution, induced by the constriction, yields an apparent double-peaked mean velocity profile in PIV analysis. Upstream of the constriction, \textit{e.g.} $x=-10$~mm in Fig.~\ref{FIG6}, the suspensions exhibits a plug-like velocity profile, similar to the results shown in Fig.~\ref{FIG4}(a), obtained through particle tracking. Passing through the narrow constriction, RBCs are accelerated rapidly and a jet flow is observed downstream of the contraction, \textit{e.g.} $x=0.25$~mm, which corresponds to roughly $0.625~W$, also clearly visible in Fig.~S1, ESI$\dag$. This jet flow transitions further downstream, \textit{e.g.} $x=0.5$~mm ($1.25~W$), into the apparent double-peaked mean velocity profile. With increasing distance from the constriction, we observe a minor decrease of magnitude of the peak velocity at $y\approx\pm150~\mu$m, while the velocity in the channel center part $-100\leq y \leq 100~\mu$m increases slightly. The non-monotonic profile persists even after $x=10$~mm post-contraction, close to the channel outlet.

Recently, Zhou~\etal\cite{Zhou2019} investigated the spatio-temporal dynamics of dilute RBC suspensions in a straight microfluidic channel at different $x$-positions. Tracking cells along the channel width, the authors reported off-center two-peak (OCTP) velocity profiles at low $\text{Re}_{\text{p}}\ll 1$ that persisted even after $20\times D_\text{h}$ and attributed the occurrence of such a heterogeneity to non-inertial hydrodynamic lift forces. Although they did not resolve the spatial distribution of cells across the channel height experimentally, complementary numerical simulations showed that RBCs concentrate at four positions close to the channel walls, similar to the results shown here.

In this study, we first proved the existence of a non-uniform spatial distribution of RBCs by tracking individual cells in different layers in the channel cross-section. This allowed us to identify the double-peaked profile of the mean velocity both in PTV as well as in PIV experiments as direct consequence of RBC focusing post-contraction. However, the occurrence of such inhomogeneities of the cross-sectional cell or tracer particle distribution, particularly in deep channels, might not always be straightforward \textit{a priori}. In that case, density-weighted averaging, \textit{i.e.} performing standard PIV analysis, over a large part of the channel cross-section can lead to misinterpretation of velocity data. Therefore, we hope that besides the detailed investigation of RBC flow properties, this study will also help to provide a better understanding about the influence of geometry-induced focusing effects on commonly employed velocimetry techniques.

\subsection{Trapping of rigid RBCs in downstream vortices}
\begin{figure}[h]
\centering
  \includegraphics[width=8.3cm]{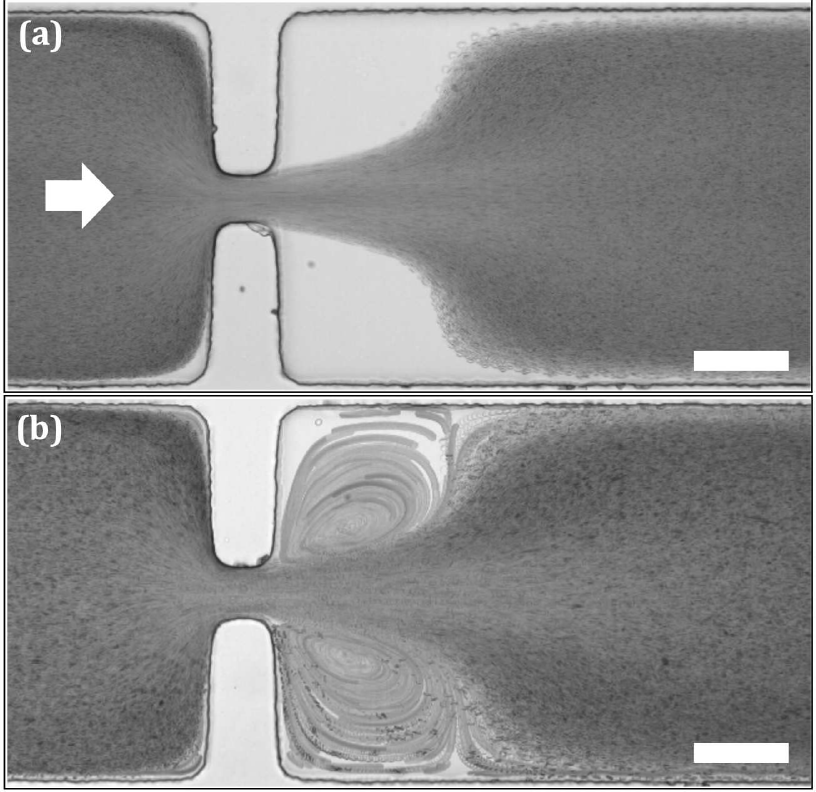}
  \caption{Time stacks of 1,050 microscopic images of (a) a $5\%~Ht$ living RBC suspension and (b) a $5\%~Ht$ rigid RBC suspension flowing through the microfluidic contraction-expansion channel at $\Delta p=240$~mbar. The direction of flow is from left to right, as indicated by the white arrow. The white scale bars represent a length of $100~\mu$m.}
  \label{FIG7}
\end{figure}

For both living and rigid RBCs, we observe similar spatial distributions and velocity profiles post-contraction, as discussed above. However, a dramatic difference arises in the vortex regions downstream of the sudden expansion, as exemplified in Fig.~\ref{FIG7} for (a) living and (b) rigid RBCs. While for living cells, the recirculation zones downstream of the constriction are completely cell-free up to the highest pressure drop ($\Delta p=240$~mbar) investigated here, rigid RBCs accumulate in the microvortices post-contraction. The size of these vortex regions, where stiffened RBCs are trapped, increases with increasing pressure drop, as shown in Fig.~S8 (ESI$\dag$). Furthermore, the cell-free layer close to the channels wall is larger for living RBCs than for rigid cells, in accordance with our previous observations (Fig.~S3, ESI$\dag$).

This phenomenon emerges as a consequence of cell deformability, which can be used for deformability-selective cell separation, \textit{e.g.} of malaria-infected RBCs.\cite{Hou2010} The effect of deformability on the microscale flow behavior of RBC suspensions has been studied previously.\cite{Passos2019, Karimi2013, Hur2011} Hur~\etal\cite{Hou2010} used an gradually expanding microfluidic chip for classification of different types of cells based on their size and deformability. They showed that flexible cells occupy equilibrium positions much closer to the channel center compared to rigid cells and used this effect to conduct label-free cancer cell enrichment. Furthermore, Yang~\etal\cite{Yang2012} used a similar device and demonstrated that rigid RBCs flow closer to the channel border, when passing the expanding part of flow channel, while living RBCs preferentially flow in the channel center. In the study presented here, we use a more abrupt expansion. We hypothesize that while living cells are deformed in the constriction and follow streamlines post-contraction, the impaired deformability of rigid cells results in an enhanced collision induced drift. This allows some of the rigid RBCs to cross streamlines sufficiently in the expanding region of the channel and to be trapped in the vortex regions post-contraction.\cite{HenriquezRivera2016} However, most of the cells pass the center of the constriction and do not linger in the recirculation zones but are focused towards the four channel face as discussed above. A detailed investigation of this effect is beyond the scope of this study but future work will include more detailed investigations of the vortex trapping of non-flexible cells. Therefore, mixtures of living and rigid cells and a more sophisticated flow channel will be used and the effect of applied pressure drop or flow rate on cell trapping will be examined for deformability-selective cell separation in the vortex regions.
%%%%%%%%%%%%%%%%%%%%%%%%%%%%%%%%%%%%%%%%%%%%%%%%%%%%%%%%%%%%%%%%%%%%%%%%%%%%%%%%%%%%%%%%%%%%%%%%%%%%%%%%%%%%%%

\section{Conclusions}
In this study, we examine the effect of an abrupt constriction on the three-dimensional distribution of RBCs in the cross-section of a rectangular microfluidic channel. While previous investigations mostly probed particle focusing in constricted flows in a two-dimensional fashion, we here reveal the cross-sectional nature of geometry-induced focusing by tracking individual cells in multiple layers across the channel depth at various $x$-positions along the flow direction of the channels.

\begin{figure}[h]
\centering
  \includegraphics[width=8.3cm]{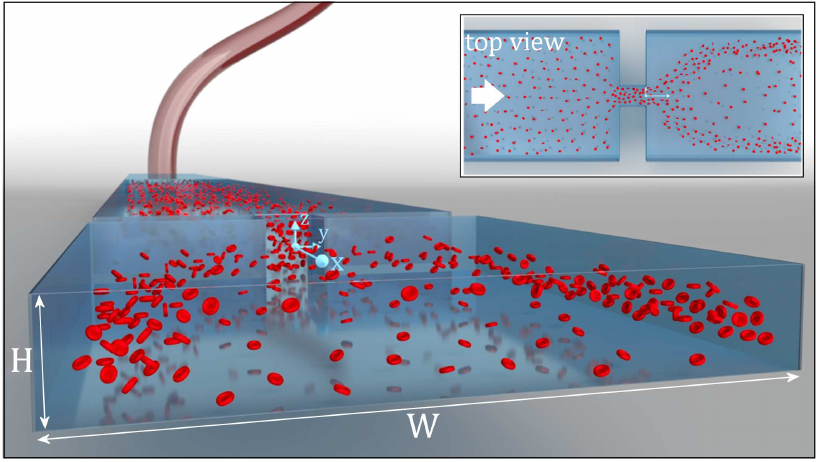}
  \caption{Schematic representation of RBCs (red) flowing through the contraction-expansion microfluidic device. Upstream of the contraction, cells are distributed homogeneously. Passing the contraction, RBCs are mainly focused in two lines near the shorter faces in the channel center plane ($z=0$ and $y/W\approx\pm 0.4$) and at the top and bottom of the channel near the walls ($z/H\approx\pm 0.3$ and $-0.4\leq y/W\leq 0.4$). In this schematic, the top channel wall is hidden and a small $Ht$ is chosen for better cell visibility. The inset represents the top view of the constriction.}
  \label{FIG8}
\end{figure}

This technique allows us to answer the specific questions stated in the introduction; (a) While the distribution of cells upstream of the constriction is uniform in the channel cross-section, without any focusing due to inertial lift forces, the sudden constriction induces a strong focusing downstream of the contraction. Here, RBCs accumulate at the top and bottom in the channel center ($z/H\approx\pm 0.3$ and $-0.4\leq y/W\leq 0.4$), as well as close to the borders of the middle plane ($z=0$ and $y/W\approx\pm 0.4$), schematically shown in Fig.~\ref{FIG8} and in accordance with numerical predictions\cite{Wang2013} on similar microfluidic constricted flows. (b) We find this non-uniform cross-sectional distribution post-contraction for both rigid and living RBCs. Covering a broad range of pressure drops $\Delta p$, we show that the magnitude of this phenomenon increases with increasing pressure drop and channel Reynolds number independent of cell deformability. However, while flexible living RBCs follow the streamlines post-contraction, some rigid RBCs are trapped in the vortex regions. (c) The observed non-uniform cell distribution leads to an apparent double-peaked profile of the mean velocity using density-weighted averaging, both in PTV as well as in PIV measurements. This non-monotonic velocity profile persists over more than $25\times$ the channel width $W$ downstream of the contraction. However, this effect arises solely as a consequence of particle focusing in combination with averaging across the whole channel depth. Therefore, this study also aims to provide a better understanding about geometry-induced focusing effects on velocimetry techniques, relevant for many different topical areas of physics, chemistry, and biology.

%%%%%%%%%%%%%%%%%%%%%%%%%%%%%%%%%%%%%%%%%%%%%%%%%%%%%%%%%%%%%%%%%%%%%%%%%%%%%%%%%%%%%%%%%%%%%%%%%%%%%%%%%%%%%%

\section*{Conflicts of interest}
There are no conflicts to declare.

\section*{Acknowledgements}
The research leading to these results has received funding from the Deutsche Forschungsgemeinschaft DFG FOR 2688 WA 1336/13-1, from the European Framework `Horizon 2020' under grant agreement number 675115 (RELEVANCE) and from the Volkswagen Foundation (Az: 93839).

%%%END OF MAIN TEXT%%%


%\balance


%%%REFERENCES%%%
\bibliography{Main} %You need to replace "rsc" on this line with the name of your .bib file
\bibliographystyle{Main} %the RSC's .bst file

\end{document}